# Pipelines for Procedural Information Extraction from Scientific Literature: Towards Recipes using Machine Learning and Data Science

Huichen Yang[*a], Carlos A. Aguirre[†a], Maria F. De La Torre[†a], Derek Christensen[†a], Luis Bobadilla[‡a], Emily Davich[‡a], Jordan Roth[‡a], Lei Luo[‡a], Yihong Theis[‡a], Alice Lam[‡a], T. Yong-Jin Han[‡b], David Buttler[‡b], William H. Hsu[†a]

[a]*Department of Computer Science Kansas State University Manhattan, Kansas, USA*
{huichen[*], caguirre97[†], marifer2097[†], derekc[†], happystep[‡], edavich[‡], jordan97[‡], leiluoray[‡], yihong[‡], lamalice16[‡], bhsu[†]}@ksu.edu
[b]*Lawrence Livermore National Laboratory, Livermore, USA*
{han5[‡], buttler1[‡]}@llnl.gov

*Abstract*—This paper describes a machine learning and data science pipeline for structured information extraction from documents, implemented as a suite of open-source tools and extensions to existing tools. It centers around a methodology for extracting procedural information in the form of *recipes*, stepwise procedures for creating an artifact (in this case synthesizing a nanomaterial), from published scientific literature. From our overall goal of producing recipes from free text, we derive the technical objectives of a system consisting of pipeline stages: document acquisition and filtering, payload extraction, recipe step extraction as a relationship extraction task, recipe assembly, and presentation through an information retrieval interface with question answering (QA) functionality. This system meets computational information and knowledge management (CIKM) requirements of metadata-driven payload extraction, named entity extraction, and relationship extraction from text. Functional contributions described in this paper include semi-supervised machine learning methods for PDF filtering and payload extraction tasks, followed by structured extraction and data transformation tasks beginning with section extraction, recipe steps as information tuples, and finally assembled recipes. Measurable objective criteria for extraction quality include precision and recall of recipe steps, ordering constraints, and QA accuracy, precision, and recall. Results, key novel contributions, and significant open problems derived from this work center around the attribution of these holistic quality measures to specific machine learning and inference stages of the pipeline, each with their performance measures. The desired recipes contain identified preconditions, material inputs, and operations, and constitute the overall output generated by our computational information and knowledge management (CIKM) system. Within the overall pipeline, we have applied machine learning approaches to step classification and, in continuing work, are applying these approaches to the subtasks of feature extraction, document filtering and classification, text payload extraction, recipe step identification, and multi-step assembly.

*Keywords*-Information Extraction, Machine Learning, Natural Language Processing (NLP), Text Analysis, Metadata

## I. Introduction

We present a machine learning-driven document analysis pipeline for scientific literature that is designed to address challenges to automation of payload extraction and identification of recipe steps using natural language processing (NLP). This paper focuses on payload filtering and extraction in Portable Document Format (PDF) files, the most common format for sharing and dissemination of scientific knowledge. The overall goal of this work is to extract *recipes*, which are defined as procedural specifications in the form of sequences of steps centered around participating tagged entities and ultimately roles and operations, from scientific publications. In this paper we focus on the extraction task itself and consider each purpose and application as a use-case of document analysis.

Our pipeline is designed on principles of holistic document analysis - specifically, to use machine learning in multiple stages with shared objectives for document analysis. Each stage passes successively refined natural language and metadata features on to the next. Our study focuses on materials process engineering, with an emphasis on detecting and categorizing techniques for the synthesis of nanomaterials, an emerging research and development area. As with our prior work on this application domain, our overall goal is to extract recipes from the scientific literature, using information extraction techniques that are based on machine learning, applied to both labeled and marked-up corpora. However, there is no existing system to date that solves holistic information extraction tasks of the desired form, such as automatic compilation of full recipes, single recipe steps, or even chemical unit operations, from published scientific papers. There is a wide technical gap between the intake of published literature from source collections and the output of actionable information such as the recipe-containing sentences of an article known to be relevant. Some existing tools, such as *PDFBox*, can convert PDF documents to text files, but cannot extract useful information from those text files. [12] Conversely, other tools can help extract domain-related information from text files, but cannot filter the documents for relevance to a query, or segment and order the appropriate text payload within a PDF file. Our system narrows this gap by interfacing such tools using format and metadata standards that are shared from stage to stage of a pipeline, and providing a unified supporting framework for the algorithms and representations of all stages that is driven throughout by machine learning. The system diagram for this pipeline and framework is depicted in Figure 1. The central tasks of this paper are the extraction, classification (and automatic annotation), and federated web-based delivery of: plain text

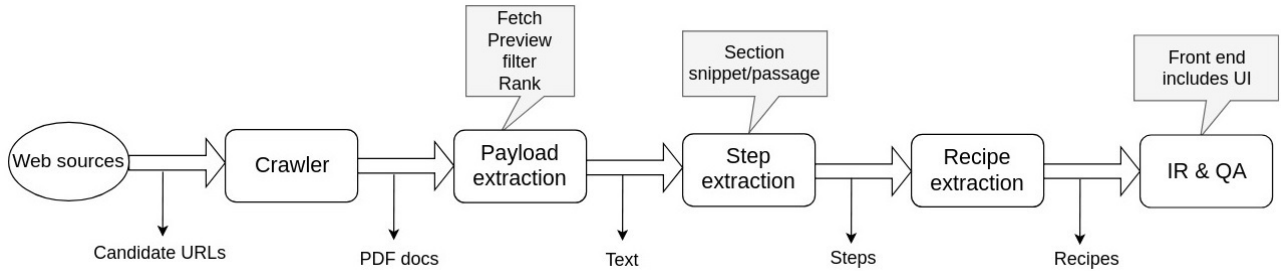

Figure 1. Extractor Pipeline

payloads, associated figures, recipe-related sentences, and finally recipe steps.

We begin by reviewing related work in Section II and in Section III, introduces the entire extractor pipeline and the methodological details for each stage. We then fully describe an experiment design and present experimental results in Section IV, and finally derive conclusions and priorities for future work in Section V.

## II. Related work

Recent work on empirical methods for NLP and supervised machine learning using extracted information has been applied to the domain of nanomaterials IE. For example, Kim et al. introduced synthesis parameters of oxide materials extraction which is based on machine learning and NLP from relevant journal articles [1]. They used an existing application programming interface (API) of CrossRef [2] to retrieve relative articles and converted these to plain text from downloaded files with HTML and PDF formats. The 'bag of words' method has been used for relevant section classification with binary logistic regression in paragraphs. These paragraphs are collected based on manual annotation from 100 different journal articles. The final text is extracted from candidate sections in a scientific article, using a combination of pre-trained Word2Vec [13] and neural network [14] models to proceed text extraction. This Word2Vec is used to learn accurate vector representations for specifying domain-related words of oxide materials. This is a first step towards using machine learning for information extraction from large and comprehensive corpora in the nanomaterials domain, and yields some methods that are potentially transferable to other domains of scientific literature. Another significant research project [3] also addresses text extraction with machine learning techniques from scientific literature in the materials synthesis domain.

## III. Information Extraction System

In this section, we describe the details of each stage of the extractor pipeline (Figure 1). Our system covers from scientific literature source collection to recipe steps displayed on web page, including the following steps:
- Synthesis literature materials are crawled from online resources. The type of this literature is free texts in basic publication formats (PDF or HTML);
- Documents (PDF) conversion to plain text and extracting relevant figures and other images from published literatures, then filter the domain-related literature;
- Step extraction/sentence classification from experiment section of literatures;
- Recipe assembly from relevant sentences;
- Front end as user interface to display the contents of whole literature which includes title, authors, abstract, body of literature, sections, references and recipe results.

### A. Crawler/Ranker/Filter

We use a crawler which automates the process by utilizing a set of seeds (e.g. URLs) to find sufficient links to look through in order to construct the corpus of our domain.

To complete the task of finding and downloading PDFs from the web, we created a Java based web crawler. To start, the crawler takes a newline delimited text file containing seeds (URLs) as the base index to start the crawl from. It then builds a B-Tree from the URLs given and the URLs crawled with a depth determined by the given depth in the configuration file. From the nodes on the B-Tree, the crawler determines which nodes are downloadable PDFs and the proceeds to download them into the output directory. This process runs until each node on the B-Tree has been covered. We also developed a backward citation component to help focus the crawler. After the first initial crawl and after the annotations of the gathered PDFs, the PDFs that were matched positive for relevancy are sent back to the crawler. The crawler then handles the metadata from the PDFs and uses the citations in each paper to query for new, more relevant, seeds. We used relevant seeds which are provided by expert and crawled 30K scientific literature from public resources as our corpus.

### B. Payload Extraction

The payload extraction system used in this work is based on similar ones developed for speeding up the annotation process to identify relevant papers from a corpus of scientific documents using classification [5]. This tool shows the first few pages of any paper, lists the keywords which are domain-related, and highlights them on the paper simultaneously. That is so users can quickly go through the paper indicated by the highlighted keywords to decide if the paper is related to target domain.

The text extraction and section classification steps are crucial for recipe extraction in the pipeline. Classifying the different sections of a synthesis literature allows for the region of interest in recipe-step search to be narrowed down. However, with the large crawled document corpora comes a disparity of document formatting and challenges for section classification. To address these format disparity challenges, *MATESC*, a tool for metadata-aware extraction developed by De La Torre et al. [6] is adapted and improved. This tool uses metadata features and heuristics, such as font size, font type and character spatial location to group words, lines and paragraphs to classify them with their corresponding section header.

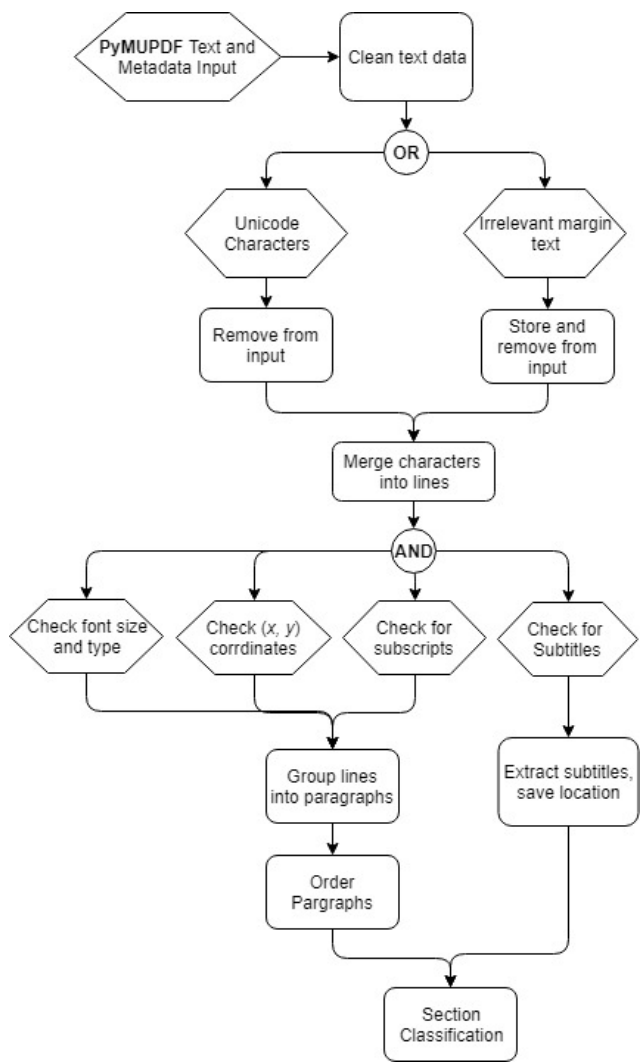

Figure 2. MATESC's Input Processing and Section Classification Pipeline

Figure 2 shows the work flow of *MATESC (Metadata-Analytic Text Extractor and Section Classifier for Scientific Publications)*, an open-source tool developed by De La Torre et al. *MATESC* takes PDF documents as input and uses PyMuPDF [7] to extract text and the metadata of each character. The extracted text is filtered by removing irrelevant text usually found in the margins of each document page, using their spatial location. Then, words are merged into their corresponding line, while considering font and spatial location to differentiate between section titles and section content. Afterwards, those lines are then grouped into paragraphs and those paragraphs are sequentially ordered based on a calculated bounding box of a paragraph. Table I shows the accuracy of *MATESC* evaluated on random articles versus articles relevant to the nanomaterials synthesis domain. For detailed results regarding section classification accuracy which we use as a baseline, we refer the interested reader to De La Torre et al. [6].

| Name | Accuracy | Precision | Recall | F1-score |
|---|---|---|---|---|
| Random Domain | 0.85 | 0.63 | 0.63 | 0.57 |
| Relevant Domain | 0.88 | 0.78 | 0.74 | 0.72 |

Table I
ACCURACY FOR THE MATESC EXTRACTOR ON RANDOM VS SYNTHESIS OF NANOMATERIAL RELEVANT PAPERS

To improve *MATESC*'s text grouping and section classification, we used the *scikit-learn* implementation of the Density-Based Spatial Clustering of Applications with Noise (DBSCAN) algorithm to classify spans of text into their corresponding groups based on the euclidean distance between each span's metadata features. The features included x,y coordinates, font type, and font size of each text span. The current limitations of the algorithm include excessive splitting and (less frequently) merging due to the unordered clustering nature of the algorithm. Some text groups-spans and lines-are over-split (under-merged) based on the difference distance threshold. Future work involves further research in DBSCAN's parameter estimation, including fine-tuning of radius and minimum point thresholds.

As our ground truth for testing *MATESC* and other clustering algorithm output, we developed an open-source dataset using *labelImg* [16] to obtain the regions of interest for headers, heading, and paragraphs. The data set allows us to use *PyMuPDF* to extract the text spans within each region of interest and create the ground truth paragraphs and sections. Figure 3 shows an example of a annotated page used as Payload Extraction ground truth. Currently, this data set allows us to test grouping, clustering and sequential reading order of these groups.

*1) Paragraph and Section Clustering:* Due to PDF document format disparity among different journals and publications, the procedure to form paragraphs and sections from characters is laborious. Heuristics that consider the metadata features provided by it PyMuPDF and others calculated from it, such as spacing between lines and column recognition, has been integrated into *MATESC*. Nevertheless, learning to cluster paragraphs and sections is an interesting problem that the group has started to work on. Grouping lines into paragraphs and paragraphs into sections is key to finding the synthesis within the literature. In the future, we plan to test clustering algorithms, such as the DBSCAN algorithm to group lines into their corresponding paragraphs based on the distance measurement of each line's metadata features. Some of these features include the bounding box around the text, the font type and size and the page number the line belongs to. Furthermore, we would like to use `pyfaster-rcnn` [15] to obtain region proposals for our sections, we hope to see an increase in the accuracy of section classification and extraction.

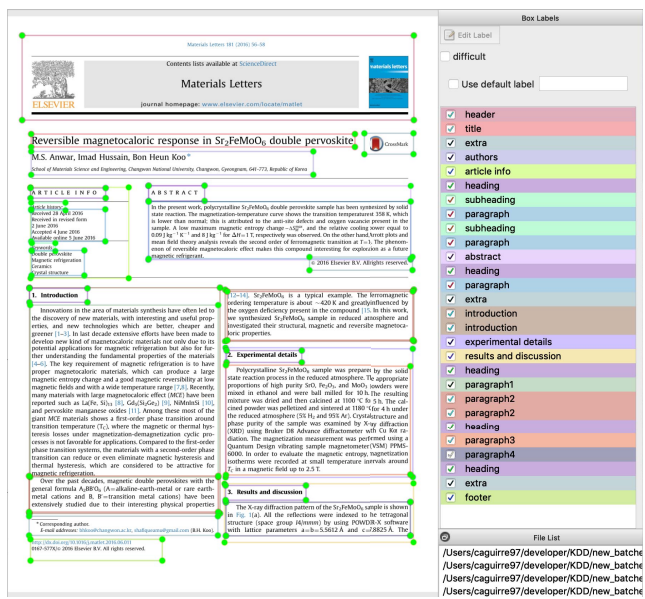

Figure 3. Payload Extraction ground truth (markup annotation)

### C. Step Extraction

We use a binary Naïve Bayes (NB) classifier to perform sentence classification on the experiment section, which was an output of the payload extraction stage. We hand-labeled 2600+ relevant or irrelevant sentences from 98 relative literatures as our training data set, $(s_1, c_x) \ldots (s_m, c_x)$ where $s$ represents a sentence, and $c$ represents its label, where the value of $c$ is 1 (relevant) or 2 (irrelevant). We then train the NB classifier to obtain a learned function $\gamma: s \rightarrow c$ that predicts the class attribute of input sentence s through the classifier.

$$c_{\text{NB}} = \underset{c \in \mathcal{C}}{\mathrm{argmax}}\, P(c_i) \prod_{s \in S} P(s|c) \qquad (1)$$

Here $S \equiv \{s_1, s_2, \ldots, s_n\}$ represents the sentences from experimental section, and $C \equiv \{c_1, c_2\}$ represents the binary classes. We also tried to train our model with different feature categories of transforming training data set to different vectors:

1) Count vectors as training features, converting a collection of sentences to a matrix of term frequency;
2) TF-IDF vectors as training features, converting a collection of sentences to a matrix of TF-IDF feature scores;
3) N-gram as training features, converting a collection of sentences to a matrix of TD-IDF scores of N-grams;

The model has been implemented in *scikit-learn* [8], and results are shown in table II. The class 0 represents irrelevant sentences, and class 1 represents relevant sentences.

| Name   | Accuracy | Class | Precision | Recall | F1-score |
|--------|----------|-------|-----------|--------|----------|
| Count  | 0.79     | 0     | 0.83      | 0.78   | 0.80     |
|        |          | 1     | 0.76      | 0.81   | 0.78     |
| TF-IDF | 0.78     | 0     | 0.81      | 0.82   | 0.82     |
|        |          | 1     | 0.77      | 0.76   | 0.76     |
| N-gram | 0.76     | 0     | 0.78      | 0.83   | 0.80     |
|        |          | 1     | 0.76      | 0.69   | 0.72     |

Table II
ACCURACY OF NB CLASSIFIER WITH DIFFERENT CATEGORIES

### D. Recipe Extraction

A *recipe* in our research is defined as a set of specific actions that are applied to a set of recognized base materials in experiments within the application domain of nanomaterials synthesis. After conducting informal elicitation sessions with subject matter experts concerning the format and content of recipes, we developed a written rubric; we then analyzed 27 relevant papers manually to extract recipes as our ground truth. *ChemicalTagger* [9], an open-source tool for semantic text-mining in the chemistry domain that is based on *OSCAR* [10] (a system for automated annotation of scientific articles in the field of chemistry or a subarea), is used in our research to extract recipes from relevant sentences generated from the stage of step extraction in our pipeline. *ChemicalTagger* generates XML files from raw text. The XML files include different tags and some of which contain the verbs labeled as action phrases (e.g. dry, wait) by *ChemicalTagger*. We then parse the XML and extract recipes from the sentences containing action phrases.

It is worth noting that we did not use *ChemicalTagger* directly but rather as a source of functional features on which our new system is based. The reason is that *ChemicalTagger* defines some verbs as action phrases which are not applicable in the nanomaterials domain. For instance, "prepare" is recognized as a verb and extracted by *ChemicalTagger*; however, it is in the stage of preparation rather than the real outcomes of recipe steps in our domain. Because of that, we have compared the results extracted by *ChemicalTagger* with our ground truth and modified the action phrases in *ChemicalTagger* to be consistent with our domain knowledge (e.g. adding injection as action phrase).

### E. Front End Intelligent UI for IR & QA

The front end interface has been developed to demonstrate our system's information retrieval capabilities. This interface presents the user with the option of querying a set of papers to view by selecting their material and morphology or the option of searching all papers by user provided search terms. Our front end shows the system's functionality after the payload extraction step has occurred. The resulting papers were initially the output of the crawler stage which were then modified for retrieval in the payload extraction stage. After a search is made, the user can view relevant paper titles with their corresponding images. If a paper is selected for viewing, a user is shown a paper's extracted contents and its DOI. A paper's extracted abstract, experimental section, and references are displayed. The capability of searching by user provided terms is implemented using the Apache Solr [11] search engine.

## IV. Recipe Evaluation

As described above, each desired recipe step output is a demarcated passage or set of passages within a sentence which includes an action followed by some specific materials and/or metrics in our domain. A complete recipe thus consists of multiple recipe stems that are combined by sequential grouping (`begin`/`end`), with the eventual goal of developing a formal specification language that includes parallel execution (`cobegin`/`coend`) and iteration. This would facilitate the development of a materials synthesis planning language beginning with a data definition language (DDL) and leading to a formal ontology for specific recipe-based tasks such as question answering (QA) about ingredients, unit operations, and embedded numerical quantities, such as concentrations and temperatures. For the recipe evaluation we need to measure the difference between the recipe output from our system and recipe ground truth, which is annotated manually. Identifying whether the recipes are the same as the recipe ground truth is a complex task that is challenging to specify formally and difficult to automate, because the recipe output also depends on each stage of the extractor pipeline (Figure 1). For example, at the Payload stage, we might lose some information, such as Unicode characters: °, ≥, or there is some extra space generated between the material names in our domain, when converting PDF document to text file. Additionally, at the Step Extraction stage, we trained machine learning model for filtering the irrelevant sentences, but the accuracy of the model would also affect the final recipe outputs. For the Recipe Extraction stage, *ChemicalTagger*, an open-source tool, was used to extract sentences that include action phrases. Whether these action phrases can fit in our domain or not would determine the accuracy of the result of recipe extraction.

By examining the holistic output at the sentence level, we calculate precision, recall. and F1 scores to evaluate the recipe output generated from our system. In our system, precision $(T_r/T_r + E_r)$ indicates the rate of recipes extracted by our system that correspond to the ground truth recipes, from a known reference set determined by annotation; recall $(T_r/T_r + M_r)$ measures the rate of known reference recipes that are successfully captured by our system; F1 score is the harmonic mean of precision and recall.

- $T_r$ (**true positives**) represents output recipes $r$ from a system that are the same as ground truth recipes;
- $E_r$ (**false positives**) represents **extra** recipes $r$ that are captured by our system but are not relevant recipes, compared with ground truth;
- $M_r$ (**false negatives**) represents $r$ that are part of the ground truth recipe but **missed** by our system;

The measurement of above parameters is based on cosine similarity, which will eventually calculate the precision, recall, and F1 score. Moreover, the recipes captured by the system are from the same document where the ground truth is located. Therefore, there is no meaning ambiguity between the two recipes in comparison and semantic analysis is not necessary.

Specifically, cosine similarity is used to evaluate how similar two documents are from each other. These two documents in the form of vectors represent the recipe output from our system and the recipe ground truth, respectively.

$$Score_{\text{similarity}} = \frac{\sum_{i=1}^{n} O_i \times G_i}{\sqrt{\sum_{i=1}^{n}(O_i)^2} \times \sqrt{\sum_{i=1}^{n}(G_i)^2}} \quad (2)$$

In Equation (2), $O$ and $G$ represent the two documents in terms of vectors, where $O$ denotes recipe output and $G$ denotes ground truth. We evaluate recipe accuracy for two perspectives: (a) measure the similarity of the two documents; (b) measure the accuracy of recipes by looping each sentence in the recipe output to compare with the recipe sentences in ground truth. Results generated from a and b will be compared and sometimes lead us to explore further. For example, a high score in terms of similarity between the two documents but a low score for accuracy of recipes is a red flag to us and further investigation is warranted. Regarding the accuracy in (b), we consider two situations in which a recipe is accurately extracted: if the similarity is equal or greater than 70%, the parameter $T_r$ would be set up to 1, meaning the truth recipe has been outputted by our system; if the similarity is greater than 50% but less than 70%, we assign the value of 0.5 to parameter $T_r$. This is because some of the recipes captured by our system actually are partial ground truth recipe. A very strict restriction will filter out them.

| Doc$_{\text{Similarity}}$ | Precision | Recall | F1 |
|---|---|---|---|
| 87.95% | 74.76% | 71.33% | 70.27% |

Table III
System Evaluation Results

Table III shows the average similarity between two documents, percentage of precision, recall and F1 score compared the system recipe output with the recipe ground truth of 27 papers which are annotated manually. The evaluation of these 27 papers has been done on a separate test set other than on the training dataset.

## V. Conclusions and Continuing Work

The procedural extraction tasks and pipeline described in this paper also represent a test bed for the document analysis tasks of learning to rank and filter in a focused crawler, text extraction from typeset documents, snippet and passage extraction, and especially unstructured to structured information extraction. Each stage of the holistic system affects the final recipe output, demonstrating a challenging problem of credit assignment that may be amenable to representations for sequential decision making. In preliminary experiments on the entire pipeline, we noticed that raising the accuracy of payload extraction, training our models with more data sets, and extracting action phrases that better fit in our domain propagate gains downstream to improve recipe extraction accuracy.

Promising findings using corpora crawled using seeds described in [5] suggests that next steps in current and future work ought to involve expansion of the test bed to other nanomaterials domains and examine transfer learning between domains (one type of material to another), and between tasks (e.g., recipe knowledge base population to recipe QA or textual entailment). The existing system incorporates information extraction in the form of full-document payloads and sentential or sub-sentential units of recipe-bearing text, and also filters collections for relevance to a specified material of interest and incorporates a presentation module for the display of extracted figures and other embedded content.

Our source code and experiments are available at https://bitbucket.org/hcyang66/icdar-ost2019.

ACKNOWLEDGMENTS

This work was funded by the Laboratory Directed Research and Development (LDRD) program at Lawrence Livermore National Laboratory (16-ERD-019). Lawrence Livermore National Laboratory is operated by Lawrence Livermore National Security, LLC, for the U.S. Department of Energy, National Nuclear Security Administration under Contract DE-AC52-07NA27344.